# The shifts X-Ray Mn Kα and 2p spectra of Mn-Heusler alloys

Shkvarin A.S., Yarmoshenko Y.M.

X-ray emission $K\alpha_{1,2}$ spectra of Mn in Heusler alloys $Co_2MnMe$ (Me = Al, Ga, Sb), $Ni_2MnIn$, $Cu_2MnAl$ were studied. Shifts of Mn $K\alpha_{1,2}$ lines relatively pure Mn in high-energy region and low-energy shifts of binding energy Mn 2p XPS is detected. X-ray emission and XPS shifts are in qualitative agreement.

The strong interest to Heusler alloys is caused by functional properties, and in particular as a half-metallic ferromagnetic, whose existence for NiMnSb was predicted by de Groot et al. in 1983 [1]. The main feature of the half-metallic magnets is the presence of an energy gap for one of the spin projections on the electron density of states. For many Heusler alloys minority spin density of states (DOS) has an energy gap in vicinity the Fermi energy, in contrast majority spin DOS has metallic character of properties as showed by the Kübler et al. [2]. Therefore, Heusler alloys are regarded as perspective candidates for application in spintronics.

The precise control at each stage of the synthesis and the final attestation of the samples is necessary in order to provide the existence of required physical properties of bulk materials, in particular for the interfaces containing layers of Heusler alloys. Full XRD often is difficult for evident reasons (e.g. insufficient bulk for analysis, necessity to destroy the sample, a crystalline texture, etc.). Photoemission (XPS and HXPS) of Mn 2p core level is sufficiently reliable probe for traditional Heusler alloys based on manganese, e.g. $Co_2MnMe$ (Me = Al, Ga, Sb) witch are crystallize with a cubic lattice $L2_1$, space group Fm3m (225) [3]. XPS Mn $2p_{3/2}$ line of Heusler alloys has a unique and characteristic for these materials energy splitting of 1 eV [4]. This spectral feature is sufficient for the attestation taking into account all the other conditions.

We have studied the X-ray $K\alpha_{1,2}$ emission spectra (XES) and Mn 2p core levels in the Heusler alloys $Co_2MnMe$ (Me = Al, Ga, Sb), $Ni_2MnIn$, $Cu_2MnAl$ (Figure 1). The Mn $K\alpha_{1,2}$ XES spectra are measured with a Johann-type X-ray spectrometer using fluorescence excitation (by Pd L x-ray radiation). A quartz crystal (plane $10 1\bar{1}$, d=3,33 Å) was used as a crystal analyser curved to R = 1.35 m. Width of the function instrumental distortion according to our estimates is 0.9 eV (E/ΔE=5900 for E=5898 eV). Registration of the secondary X-rays was performed by linear coordinate detector with gas flow (gas mixture Ar-CH4) own manufacturing with a spatial resolution of 100 micron. The accuracy of determining the energy position for the entire period of spectra measurement was 0.05 eV. For energy calibration the spectrum of pure α-Mn was measured before and after sample spectrum measurement. The photoelectron Mn 2p core level spectra of the samples were obtained with a PHI 5600 CI Multitechnique System XPS spectrometer using monochromatized Al Kα radiation. Energy resolution was of about 0.35 eV. Before measurements the samples were cleaved in the volume of the spectrometer. The XPS spectra were calibrated based on the Au 4f core level and Co 2p core level as the internal standard for cobalt-containing samples. The energy position of the Co 2p core-levels remained constant in Heusler alloys according to our data. Energy characteristics for the spectra of five Heusler alloys are shown in the Table 1. Shifts of the X-ray emission lines are given relative to the position of the spectral maximum $K\alpha_1$ line of pure manganese (α-Mn). The observed energy shift of Mn $K\alpha_{1,2}$ line qualitatively corresponds to reduction of the binding energy of Mn 2p line (Table 1, Figure 1). The divergence of the data on the magnitude of shifts for K alpha and XPS obviously is related to bulk data for XES in contrast to the surface data for $2p_{3/2}$ XPS. In [5] have been studied XES Kα and Kβ spectra in $Cu_2MnAl$. The exact energy position of the lines was not determined.

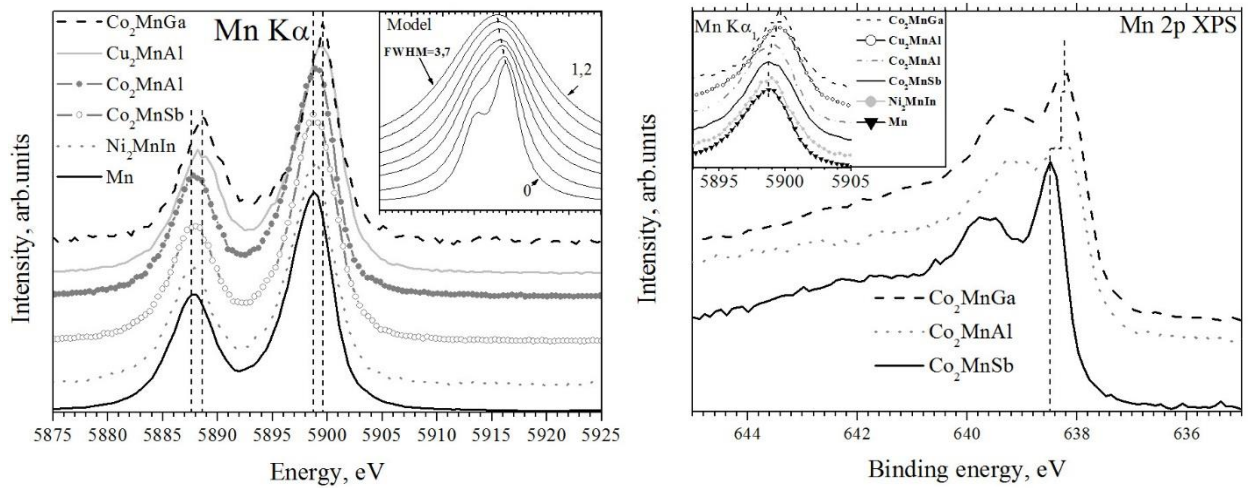

Figure 1. On left panel are shown Mn $K\alpha_{1,2}$ spectra of Heusler alloys under investigation. The inset shows the result of the spectrum modeling by convolution (see text). On right panel are shown Mn $2p_{3/2}$ spectra of samples. The inset shows Mn $K\alpha_1$ spectra for better visibility of spectral shifts.

Table 1. Energy characteristics of the spectra. XES shifts - shifts of $K\alpha_1$ line of samples relative to the α-Mn. XPS shifts - shifts of Mn $2p_{3/2}$ maximum relative to pure Mn [6]. XES FWHM - the ratios of XES FWHM of samples and α-Mn. SO (XPS) - values of spin-orbit splitting in Mn 2p XPS spectra (between Mn $2p_{3/2}$ and $2p_{1/2}$ maxima). SO (XES) - values of spin-orbit splitting in Mn $K\alpha_{1,2}$ spectra (between $K\alpha_1$ and $K\alpha_2$). $I_1/I_2$ - relation between $K\alpha_1$ and $K\alpha_2$

| Sample | XES shift | XPS shift | XES FWHM | SO (XPS) | SO (XES) | $I_1/I_2$ |
|---|---|---|---|---|---|---|
| $Co_2MnGa$ | 0.85 | -0,62 | 0,95 | 11.7 | 10.96 | 0,6 |
| $Co_2MnAl$ | 0.3 | -0,5 | 1,04 | 11.69 | 10.93 | 0,53 |
| $Co_2MnSb$ | 0.1 | -0,43 | 1,04 | 11.76 | 10.98 | 0,5 |
| $Ni_2MnIn$ | 0.05 | - | 1,11 | - | 10.99 | 0,53 |
| $Cu_2MnAl$ | 0.61 | - | 1,1 | - | 11 | 0,54 |

The most intense doublet Mn $K\alpha_{1,2}$ is the result of a dipole transition 1s → 2p in the final state $2p^53d^5s^2$. The main channel of atomic relaxation as a result of this transition is a cascade Auger transitions involving 2p-3s, 3p and valence electrons. The observed energy shift of Mn $K\alpha_{1,2}$ line corresponds to a decrease of the binding energy the main peak of Mn $2p_{3/2}$ lines (Table1). Two alternative viewpoints exists about the origin of the main line in the Mn $2p_{3/2}$ spectrum of Heusler alloys. Initially it was assumed that this form of the spectrum corresponds to the ground state with configuration $2p^53d^5s^2$ [7]. After new experimental data received for thin Mn-containing films it has been suggested another hypothesis. According to this, the shape of the spectrum due to the effect of the final state and the main line corresponds of the excited term of the atom of manganese $2p^53d^6s^2$ (see for example [8]).

Let us consider what form and the energy position will be takes Mn $K\alpha_1$ spectrum if the Mn$2p_{3/2}$ represent the initial state. The inset of Figure 1 shows a simple model, illustrating $K\alpha$ spectrum as a convolution 1s core level and splitted on 1 eV 2p level $\Lambda(1s)*P(2p_{3/2})$, where $\Lambda(1s)$ the Lorentz function. Smearing of the original spectrum leads to the successive shift of 0.3 eV towards decreasing photon energy. Our experimental Mn $K\alpha_1$ have a half-width $\Delta E = 3.7$ eV. Full value of the hardware half-width distortion function is 1 eV in comparison with the data of [9], where $\Delta E = 2.7$ eV for Mn $K\alpha_1$. Nevertheless, under these conditions, the maximum spectral line occupies an asymmetrical position with respect to the position of the spectral components. This effect caused by the asymmetry of the Mn $2p_{3/2}$ lines should lead to an increase in the energy of the spin-orbit splitting in Mn $K\alpha$. However, as can be seen from the table, SO remains constant and equal to obtained for pure manganese. This fact is an argument in favor of the satellite origin of low-energy line in the Mn $2p_{3/2}$ spectrum of the Heusler alloys.
The shift towards low-energy (and irregular shape) of Mn 2p spectrum was observed during the deposition of manganese on various substrates, such as Pd, [8]. Increase of the binding energy Mn 2p lines is observed with the increasing content of the manganese atoms on the surface. Low-energy shift of the Mn K$\beta_1$ line on the BiMnO3 was observed in conditions of increasing pressure [10].The energy shifts in both the above cases according to these works is caused a corresponding decrease in the spatial localization of Mn 3d electrons. Therefore, information about the energy shift of Mn 2p spectra in the Heusler alloys, and in particular about the different magnitude of this shift seems productive for the interpretation of their physical properties based on the analysis of spatial localization of Mn 3d electrons.

Acknowledgments
The authors thank V.E. Dolgikh for the technical support in the preparation and performance of the experiment.

References
[1] R. A. de Groot, F. M. Mueller, P. G. van Engen, and K. H. J. Buschow, Phys. Rev. Lett. 1983, **50**, 2024
[2] J. Kubler, A.R. Williams and C.B. Sommers 1983 Phys. Rev. B **28** 1745
[3] Peter J. Webster "Heusler alloys" Contemporary Physics 1969, **10,** №6
[5] Vainshtein E.E., Kotlyar B.I. Sov.Phys.Dokl, 1956, **1**, 527
[4] Yu.M. Yarmoshenko, M.I. Katsnelson, E.I. Shreder, E.Z. Kurmaev, A. Slebarski, S. Plogmann, T. Schlatholter, J. Braun, and M. Neumann Eur. Phys. J. B 1998, **2**, 1–3
[6] A.B.Mandale, S. Badrinarayanan J. Electron Spectrosc. Relat. Phemon. 1990, **53**, 87
[7] S. Plogmann, T. Schlathölter, J. Braun, M. Neumann, Yu. M. Yarmoshenko, M. V. Yablonskikh, E. I. Shreder, E. Z. Kurmaev, A. Wrona, and A. Ślebarski Phys. Rev. B 1999, **60**, 6428
[8] A. Sandell, A.J. Jaworowski, J. Electron Spectrosc. Relat. Phemon. 2004, **135,** 7–14.
[9] M.A. Blokhin Metody rentgeno-spectralnyh issledovanij (Methods of X-ray spectral studies) 1959, Moskow, 388pp.
[10] J. M. Chen, S. C. Haw, J. M. Lee, S. A. Chen, K. T. Lu, S. W. Chen, M. J. Deng, Y.-F. Liao, J. M. Lin, B. H. Chen, F. C. Chou, N. Hiraoka, H. Ishii, K. D. Tsuei, and Eugene Huang Phys. Rev. B 2012, **86**, 045103